\documentclass[apj]{emulateapj}

\usepackage{graphicx}
\usepackage{apjfonts}
\usepackage{amsmath}
\usepackage[raggedright,FIGTOPCAP]{subfigure}
\usepackage{color}
\usepackage{hyperref}

\newcommand{\JpsrA}{J0554+3107}
\newcommand{\JpsrB}{J1422$-$6138}
\newcommand{\JpsrC}{J1522$-$5735}
\newcommand{\JpsrD}{J1932+1916}

\newcommand{\psrA}{PSR~J0554+3107}
\newcommand{\psrB}{PSR~J1422$-$6138}
\newcommand{\psrC}{PSR~J1522$-$5735}
\newcommand{\psrD}{PSR~J1932+1916}

\newcommand{\snrA}{G$179.0+2.6$}
\newcommand{\snrC}{G$321.9-0.3$}

\def\EAH{\textit{Einstein@Home}}
\def\Fermi{\textit{Fermi}}

\shorttitle{Four Gamma-ray Pulsars with Einstein@Home} 
\shortauthors{\sc Pletsch et al.}

\begin{document}

\title{Einstein@Home discovery of four young gamma-ray pulsars in \textit{Fermi} LAT data} 

\author{
H.~J.~Pletsch\altaffilmark{1,2,3},
L.~Guillemot\altaffilmark{4,5,6},
B.~Allen\altaffilmark{1,7,2},
D.~Anderson\altaffilmark{8},
C.~Aulbert\altaffilmark{1,2},
O.~Bock\altaffilmark{1,2},
D.~J.~Champion\altaffilmark{4}, 
H.~B.~Eggenstein\altaffilmark{1,2},
H.~Fehrmann\altaffilmark{1,2},
D.~Hammer\altaffilmark{7},
R.~Karuppusamy\altaffilmark{4},
M.~Keith\altaffilmark{9}, 
M.~Kramer\altaffilmark{4,10},\\
B.~Machenschalk\altaffilmark{1,2},
C.~Ng\altaffilmark{4}, 
M.~A.~Papa\altaffilmark{1,7},
P.~S.~Ray\altaffilmark{11}, 
X.~Siemens\altaffilmark{7}
}
\altaffiltext{1}{Max-Planck-Institut f\"ur Gravitationsphysik (Albert-Einstein-Institut), D-30167 Hannover, Germany}
\altaffiltext{2}{Institut f\"ur Gravitationsphysik, Leibniz Universit\"at Hannover, D-30167 Hannover, Germany}
\altaffiltext{3}{email: \href{mailto:holger.pletsch@aei.mpg.de}{holger.pletsch@aei.mpg.de}}
\altaffiltext{4}{Max-Planck-Institut f\"ur Radioastronomie, Auf dem H\"ugel 69, D-53121 Bonn, Germany}
\altaffiltext{5}{email: guillemo@mpifr-bonn.mpg.de}
\altaffiltext{6}{Currently at Laboratoire de Physique et Chimie de l'Environnement et de l'Espace -- Universit\'{e} d'Orl\'{e}ans / CNRS, F-45071 Orl\'{e}ans Cedex 02, France}
\altaffiltext{7}{Department of Physics, Univ.~of Wisconsin -- Milwaukee, USA}
\altaffiltext{8}{Space Sciences Lab, Univ.~of California -- Berkeley, USA} 
\altaffiltext{9}{CSIRO Astronomy and Space Science, Australia Telescope National Facility, Australia}
\altaffiltext{10}{Jodrell Bank Centre for Astrophysics, School of Physics and Astronomy, The University of Manchester, Manchester M13 9PL, UK}
\altaffiltext{11}{Space Science Division, Naval Research Laboratory, Washington, DC 20375-5352, USA}

\begin{abstract} 
\noindent
We report the discovery of four gamma-ray pulsars,
detected in computing-intensive blind searches of data 
from the \Fermi{} Large Area Telescope~(LAT). 
The pulsars were found using a novel search approach,
combining volunteer distributed computing via \EAH{}
and methods originally developed in gravitational-wave astronomy.
The pulsars PSRs~\JpsrA, \JpsrB, \JpsrC, and \JpsrD{} 
are young and energetic, with characteristic ages between $35$ and $56$\,kyr and  
spin-down powers in the range $6\times10^{34}$ -- $10^{36}$\,erg\,s$^{-1}$.
They are located in the Galactic plane and have rotation rates of less than $10$\,Hz,
among which the $2.1$\,Hz spin frequency of \psrA{} is the slowest of any known 
gamma-ray pulsar.
For two of the new pulsars, we find supernova remnants coincident
on the sky and discuss the plausibility of such associations.
Deep radio follow-up observations found no pulsations, 
suggesting that all four pulsars are radio-quiet as viewed from Earth. 
These discoveries, the first gamma-ray pulsars found by volunteer computing, 
motivate continued blind pulsar searches of the many other unidentified LAT 
gamma-ray sources.
\end{abstract}


\section{Introduction}\label{s:intro}

The Large Area Telescope \citep[LAT;][]{generalfermilatref} 
aboard the {\em Fermi Gamma-ray Space Telescope} 
has established pulsars 
as predominant Galactic gamma-ray sources at GeV energies \citep{Fermi2PC}. 
Most LAT-detected gamma-ray pulsars 
have been unveiled \emph{indirectly}. In these cases,
known radio pulsar ephemerides are used to
assign rotational phases to gamma-ray photons and probe for pulsations.
Additionally, dedicated radio searches at 
positions of unidentified gamma-ray sources 
as in the \Fermi-LAT Second Source 
Catalog \citep[2FGL;][]{FermiSecondSourceCatalog}
discovered many new radio pulsars, likewise 
providing ephemerides for gamma-ray phase-folding \citep[e.g.,][]{Guillemot2012,Fermi2PC}.

For the first time, pulsars have also been detected in \emph{direct} searches 
for periodicity in the sparse LAT gamma-ray photons \citep{16gammapuls2009}. 
In fact, many pulsars found in such ``blind'' searches 
are undetected at radio wavelengths \citep{Ray+2012,Fermi2PC}.
Blind searches are computationally challenging, because the relevant pulsar 
parameters are unknown in advance \citep[e.g.,][]{Chandler2001}. 
The challenge is to search a dense grid 
covering a multidimensional parameter space 
(for isolated systems: sky location, frequency~$f$, and spin-down rate~$\dot f$).
The number of grid points to be individually tested increases rapidly with 
coherent integration time \citep[e.g.,][]{bccs1:1998}: for observations spanning multiple years 
the finite computing power available makes blind searches 
with fully coherent (``brute-force'') methods unfeasible, 
and much more {\em efficient} methods are essential.

During the first year of the \Fermi{} mission pioneering blind searches revealed
24 pulsars in LAT data \citep{16gammapuls2009,8gammapuls2010} 
through a clever time-differencing technique \citep{Atwood2006,Ziegler2008}
that exploits that sparsity of the LAT data.
Increasing data time spans intensify the computing burden,
and only two more pulsars were found in the second year \citep{Fermi2PC}.
However, hundreds of unidentified LAT sources with pulsar-like properties 
\citep{2012ApJ...753...83A,Lee+2012} probably harbor undiscovered pulsars.

The blind-search problem is analogous to searches for continuous 
gravitational waves (GWs) emitted from spinning neutron stars \citep{bccs1:1998}, 
also called ``GW pulsars''. This similarity has motivated us to use 
data-analysis methods originally developed for GW-pulsar detection 
\citep[][]{bc2:2000,cutler:2005,PletschAllen2009,Pletsch2010,PletschSLCW2011} 
to significantly enhance the sensitivity of blind searches for gamma-ray pulsars.

Using these methods to search LAT data 
has led to the discovery of $10$ new gamma-ray pulsars
\citep{Pletsch+2012-9pulsars,Pletsch+2012-J1838} on the \textit{Atlas} computing 
cluster in Hannover. 
While these discoveries were isolated young pulsars 
(with spin frequencies of $3$ - $12$\,Hz), 
the ongoing searches also cover the higher-frequency range of millisecond pulsars (MSPs). 
With partial orbital constraints from optical data \citep[][]{Romani2012}, these methods also
discovered a binary MSP via gamma-ray pulsations \citep{Pletsch+2012-J1311}.

Searching for fast spinning isolated MSPs 
dominates the overall computing cost of this survey.
To enlarge the available computational resources we have recently
moved the survey onto the volunteer computing system \EAH\footnote{
\href{http://einstein.phys.uwm.edu/}{http://einstein.phys.uwm.edu/}}.
Here, we present the \EAH{} discovery and key parameters of four young, energetic
pulsars,  \psrA, \psrB, \psrC, and \psrD, detected in the ongoing blind survey of 
unidentified LAT sources. These are the first gamma-ray pulsars discovered using 
volunteer computing.

\section{Survey and Pulsar Discoveries}\label{s:disco}

The survey targets unidentified 2FGL sources with properties reminiscent 
of known pulsars. Such selection criteria include significantly curved emission 
spectra and low flux variability over time \citep{2012ApJ...753...83A}, 
leading to a list of 109 2FGL sources \citep{Pletsch+2012-9pulsars}.
Further details on source selection and data preparation 
are described in \citet{Pletsch+2012-9pulsars}.

\begin{figure}
\centerline{
\hfill
\includegraphics[width=\columnwidth]{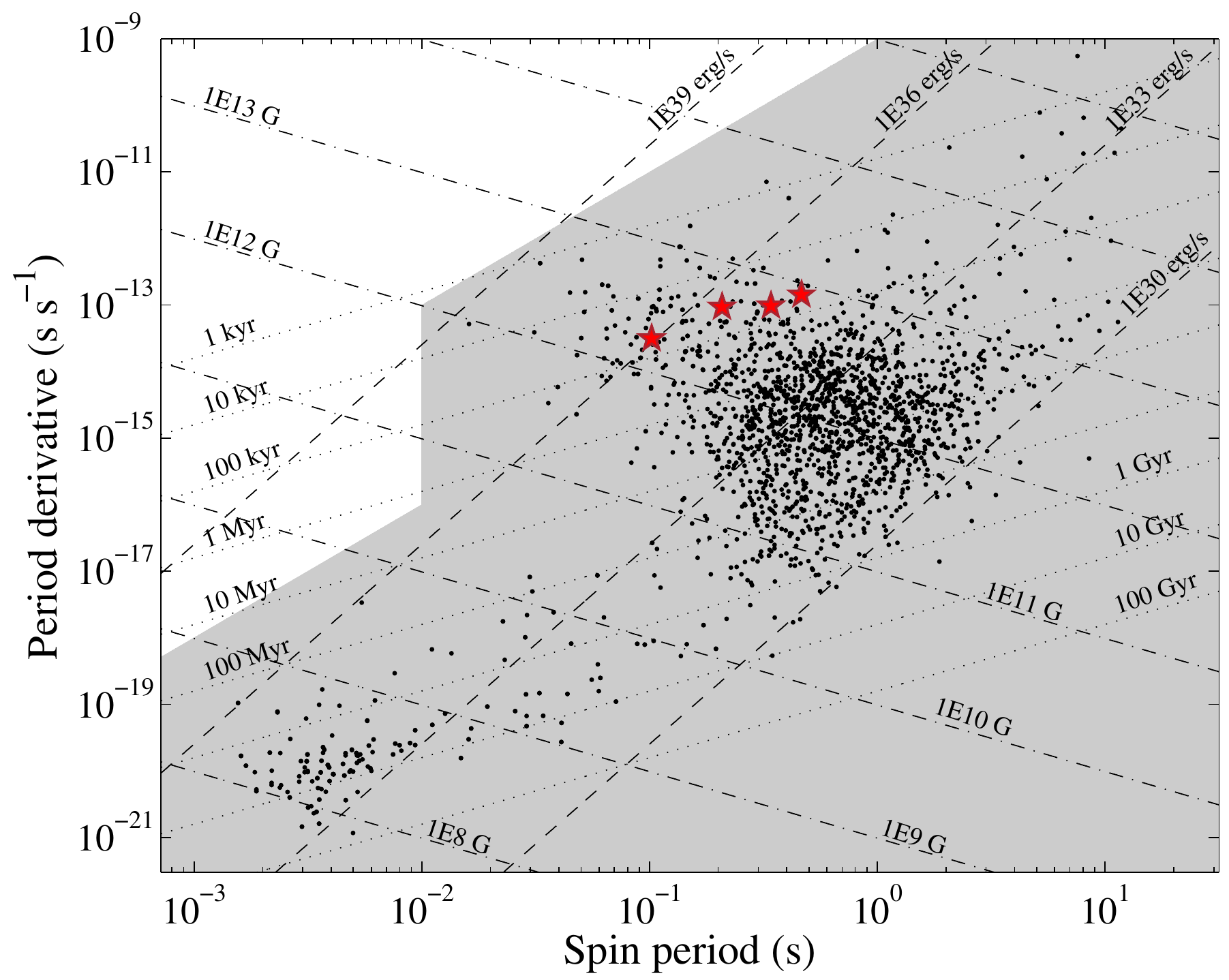}
\hfill
}
\caption{\label{f:f-fdot-diagr} 
Period -- period derivative diagram of the pulsar population. 
Filled stars show the four new gamma-ray pulsars.
Black dots represent known radio pulsars from the ATNF 
catalog \citep{ATNFcat}.
The gray shaded region indicates the search parameter space. 
Dotted lines indicate contours of constant~$\tau_c$;
similarly, dashed lines show spin-down power~$\dot E$
and dashed-dotted lines refer to surface magnetic field strength~$B_{S}$.
}
\end{figure}

For each selected target source, a blind search
for isolated gamma-ray pulsars has been carried out in three years of LAT data.
The parameter space of the search is four-dimensional (sky position, 
spin-frequency $f$, and $\dot f$). In the sky, a circular region is searched, 
centered on the 2FGL-catalog source location having a radius 20\% 
larger than the semi-major axis of the 95\% confidence elliptical error region. 
The survey covers an $f$ range up to $1.4$\,kHz, 
in order to be sensitive to MSPs. 
For $f  \geq 100$\,Hz, spin-down rates in the range
\mbox{$-$1$\times$10$^{-12}$ Hz s$^{-1} \leq \dot f \leq 0$} are searched (see Figure~\ref{f:f-fdot-diagr}). 
To maintain sensitivity to young pulsars, for $f < 100$\,Hz the $\dot f$ search 
range is extended down to characteristic ages $\tau_c =-f/2\dot f \sim 1$\,kyr, 
comparable to that of the Crab pulsar.  

The data-analysis strategy employed in the blind search follows
a hierarchical (multistage) approach, outlined in \citet{Pletsch+2012-9pulsars}.
The first stage explores the full parameter space with an efficient semi-coherent method: 
coherent Fourier powers computed with a $\approx$6~day window are 
incoherently summed as the window slides along the entire data set. 
A parameter-space 
metric  \citep{PletschAllen2009,Pletsch2010,Pletsch+2012-9pulsars} is used 
to build an efficient search grid. 
In the second stage, only significant semi-coherent candidates are 
followed up via a fully coherent analysis. A third stage further refines coherent 
pulsar candidates by using higher 
signal harmonics \citep[adopting the $H$-test of][]{deJaeger1989}. 

The \EAH{} volunteer supercomputer does the bulk of the computational work.
\EAH{} was launched in 2005 to search for GW pulsars in detector data from 
the \emph{LIGO}-\emph{Virgo} Collaboration \citep{S4EAH,S5R1EAH,S5R5EAH}.
Since 2009, \EAH{} has also been analyzing radio telescope data,  
finding several new radio pulsars \citep{EahRadiopulsar1, 2011ApJ...732L...1K, 
2013ApJ...774...93K,Allen+2013}. In parallel, \EAH{} is now also searching for
gamma-ray pulsars as described here.  This extends the radio and GW efforts
with a third distinct search for new neutron stars.

To sign up for \EAH{}, members of the general public download free
software for their Windows, Apple, or Linux computers or Android
device. Working in the background, the software
automatically downloads work units (executables and data) from the
\EAH{} servers, carries out a search when the host machine is idle,
and reports back results.  Returned results are automatically
validated by comparison of the outcome for the same work unit produced
by a different volunteer's host.  With more than 300,000 individuals
already contributing, the sustained computing capacity achieved
($1$\,PFlop/s) is comparable with the world's largest supercomputers.

The \EAH{} results 
for the formerly unidentified LAT sources, 2FGL~J0553.9+3104, 
2FGL~J1422.5-6137c, 2FGL~J1521.8-5735, and 2FGL~J1932.1+1913, 
indicated significant pulsations.
All but one of these sources also have counterparts in the
\Fermi{} LAT First Source Catalog~\citep[1FGL;][]{FermiFirstSourceCatalog},
denoted by  1FGL~J0553.9+3105, 1FGL~J1521.8-5734c, and 
1FGL~J1932.1+1914c.
A dedicated follow-up investigation to further refine the parameters and 
properties of the newly discovered pulsars is described below.

\begin{deluxetable*}{lllll}
\tablewidth{\textwidth}
\tablecaption{\label{t:parameters} Measured and Derived Pulsar Parameters}
\tablecolumns{2}
\tablehead{
\colhead{Parameter} & \colhead{\psrA} & \colhead{\psrB} & \colhead{\psrC} & \colhead{\psrD}
}
\startdata
Right ascension, $\alpha$ (J2000.0)\dotfill 
& $05^{\rm h}54^{\rm m}05\fs01(3)$ 
& $14^{\rm h}22^{\rm m}27\fs07(1)$ 
& $15^{\rm h}22^{\rm m}05\fs3(1)$ 
& $19^{\rm h}32^{\rm m}19\fs70(4)$\\
Declination, $\delta$ (J2000.0)\dotfill   
& $+31\arcdeg07\arcmin41\arcsec(4)$
& $-61\arcdeg38\arcmin28\arcsec(1)$
& $-57\arcdeg35\arcmin00\arcsec(1)$
& $+19\arcdeg16\arcmin39\arcsec(1)$\\
Galactic longitude, $l$ (\arcdeg)\dotfill   
& $179.1$ 
& $313.5$ 
& $322.1$ 
& $54.7$ \\
Galactic latitude, $b$ (\arcdeg)\dotfill
& $2.70$ 
& $-0.66$
& $-0.42$
& $0.08$\\
Spin frequency, $f$ (Hz)\dotfill 
&    $2.15071817570(7)$ 
&    $2.932827817(1)$
&    $9.790868913(3)$
&    $4.8027301667(3)$ 
\\
Frequency 1st derivative, $\dot f$ ($10^{-12}$ Hz s$^{-1}$)\dotfill 
& $-0.659622(5)$
& $-0.83293(9)$ 
& $-2.9946(2)$
& $-2.14916(1)$  
\\
Frequency 2nd derivative\tablenotemark{a}, $\ddot f$ ($10^{-23}$ Hz s$^{-2}$)\dotfill 
& $0.18(2)$
& $-1.5(3)$
& $-2.6(5)$
& $-0.3(1)$ \\
Epoch (MJD)\dotfill  
& $55214$ 
& $55214$ 
& $55250$ 
& $55214$  \\
Weighted $H$-test (single-trial false alarm probability) \dotfill 
& $425$  $(\sim 10^{-80})$ 
& $469$  $(\sim 10^{-89})$
& $319$  $(\sim 10^{-59})$
& $460$  $(\sim 10^{-87})$ \\
\\
Epoch of glitch 1 (MJD)\dotfill
& - 
& $55310$
& $55250$
& -\\
Permanent $f$ increment, $\Delta f_{\rm glitch\,1}$ ($10^{-6}$ Hz)\dotfill 
& -
& $0.026528575(2)$
& $-0.112(6)$
& -  \\
Permanent $\dot f$ increment, $\Delta \dot{f}_{\rm glitch\,1}$ ($10^{-15}$ Hz s$^{-1}$)\dotfill 
& -
& $6.4(4.8)$
& $3.6(4)$
& -  \\
Decaying $f$ increment, $\Delta f_{\rm d, glitch\,1}$ ($10^{-6}$ Hz)\dotfill 
& -
& -
& $0.42(6)$
& -  \\
Decay time constant, $\tau_{\rm d, glitch\,1}$ (days)\dotfill 
& -
& -
& $27(5)$
& -  \\
\\
Epoch of glitch 2 (MJD)\dotfill
& - 
& $55450$
& -
& -\\
Permanent $f$ increment, $\Delta f_{\rm glitch\,2}$ ($10^{-6}$ Hz)\dotfill 
& -
& $1.18798307(4)$
& -
& -  \\
Permanent $\dot f$ increment, $\Delta \dot{f}_{\rm glitch\,2}$ ($10^{-15}$ Hz s$^{-1}$)\dotfill 
& -
& $-5.3(5.0)$
& -
& -  \\
\\
Characteristic age, $\tau_c$ (kyr)\dotfill 
& $51.7$
& $55.8$
& $51.8$
& $35.4$    \\
Spin-down power, $\dot E$ ($10^{34} {\rm erg\,s^{-1}}$)\dotfill 
&    $5.6$
&    $9.6$
&    $115.7$
&     $40.7$ \\
Surface magnetic field strength, $B_{\textrm{S}}$ ($10^{12}$G)\dotfill
&     $8.2$
&     $5.8$ 
&     $1.8$ 
&     $4.5$ \\
Light-cylinder magnetic field strength, $B_{\textrm{LC}}$ (kG)\dotfill 
&     $0.8$
&     $1.3$ 
&     $15.6$ 
&     $4.5$ \\
Estimated maximum distance\tablenotemark{b}, $d_{100\%}$  (kpc)\dotfill 
&   $\lesssim 5.2$
&   $\lesssim 4.8$
&   $\lesssim 12.5$
&   $\lesssim 6.6$\\
Estimated ``heuristic'' distance\tablenotemark{c}, $d_{h}$  (kpc)\dotfill 
&   $\sim 1.9$
&   $\sim 1.5$
&   $\sim 2.1$
&   $\sim 1.5$\\
\\
Spectral index, $\Gamma$ \dotfill
&     $1.1 \pm 0.2$
&     $0.3 \pm 0.2$ 
&     $1.4 \pm 0.2$ 
&     $1.7 \pm 0.1$ \\
Cutoff energy, $E_c$ (GeV) \dotfill
&     $1.3 \pm 0.2$
&     $2.5 \pm 0.3$
&     $1.5 \pm 0.3$
&     $1.2 \pm 0.2$ \\
Photon flux above 100\,MeV, $F_{100}$ ($10^{-8}$ photons cm$^{-2}$ s$^{-1}$)\dotfill
&     $1.9 \pm 0.3$
&     $1.1 \pm 0.2$
&     $8.5 \pm 1.1$
&     $15.1 \pm 1.0$ \\
Energy flux above 100\,MeV, $G_{100}$ ($10^{-11}$ erg cm$^{-2}$ s$^{-1}$) \dotfill
&     $1.7 \pm 0.1$
&     $3.5 \pm 0.3$
&     $6.2 \pm 0.4$
&     $7.8 \pm 0.4$\\
\\
Peak multiplicity \dotfill
& $3$
& $2$
& $1$
& $1$ \\
FWHM$_{\mbox{\tiny Peak $1$}}$ \dotfill
& $0.10 \pm 0.02$ 
& $0.12 \pm 0.02$ 
& $0.22 \pm 0.02$ 
& $0.23 \pm 0.03$ 
\\
FWHM$_{\mbox{\tiny Peak $2$}}$ \dotfill
& $0.06 \pm 0.02$ 
& $0.11 \pm 0.01$ 
& - 
& - 
\\
FWHM$_{\mbox{\tiny Peak $3$}}$ \dotfill
& $0.03 \pm 0.01$ 
& - 
& - 
& - 
\\
Peak $1$ to $2$ separation \dotfill
& $0.24 \pm 0.01$ 
& $0.18 \pm 0.01$ 
& - 
& - 
\\
Peak $1$ to $3$ separation \dotfill
& $0.35 \pm 0.01$ 
& - 
& - 
& - \\
\\
Radio-flux-density upper limit at 1.4\,GHz, $S_{1400}$ ($\mu$Jy) \dotfill
& 66 
& 60 
& 34 
& 75 \\
\enddata
\tablecomments{The data time span is $54702$ -- $56383$\,MJD. 
The JPL DE405 Solar System ephemeris has been used; times refer to Barycentric Dynamical Time (TDB). 
Numbers in parentheses are statistical 1$\sigma$ errors in the last digits.
\tablenotetext{a}{Parameterizes timing noise 
(and glitch recovery where applicable) rather than pulsar intrinsic spin-down.}
\tablenotetext{b}{Assuming 100\% efficiency ($L_\gamma = \dot E$) and $f_\Omega = 1$, 
gives rise to $d_{100\%} = (\dot E / 4\pi\,G_{100})^{1/2}$.}
\tablenotetext{c}{Assuming a ``heuristic'' luminosity \mbox{$L_\gamma^h = (\dot E/10^{33}$\,erg s$^{-1})^{1/2}$\,$10^{33}$\,erg s$^{-1}$}  and $f_\Omega = 1$, 
yields $d_{h} = (L_\gamma^h / 4\pi\,G_{100})^{1/2}$.}
}
\end{deluxetable*}

\section{The Four Gamma-ray Pulsars}\label{s:fu-analysis}

For follow-up analysis, we extended the original data sets to include LAT photons 
recorded from 2008 August~4 until 2013 April~1. 
We used the \Fermi{} Science 
Tools\footnote{\href{http://fermi.gsfc.nasa.gov/ssc/data/analysis/scitools/overview.html}{http://fermi.gsfc.nasa.gov/ssc/data/analysis/scitools/overview.html}} (STs)
to select ``Source''-class photons according to the P7\_V6 instrument response functions (IRFs),
with reconstructed directions within 15\degr of the pulsars, 
energies above 100 MeV, and zenith angles $\leq 100$\degr.
We excluded photons recorded when the LAT's rocking angle exceeded 52\degr, 
or when the LAT was not in nominal science mode.
We assigned each photon a weight measuring the probability of having
originated from the pulsar (as was also done for the search). 
These weights \citep{KerrWeightedH2011} were computed with \textit{gtsrcprob} based on 
a spectral model of the region (described below) and the LAT IRFs. This photon-weighting scheme
improves the signal-to-noise ratio of the pulsations, providing better background
rejection than simple angular and energy cuts.

With these LAT data sets, we refined the initial pulsar parameters after discovery,
using the methods by \citet{Ray2011}. We subdivided the data sets 
into segments of about equal length and produced pulse profiles for all 
segments by folding the photon times with the initial parameters. 
These pulse profiles were correlated with ``template'' profiles to obtain 
pulse times of arrival (TOAs). Using \textsc{Tempo2} \citep{Tempo2} we fitted
the TOAs to a timing model with sky position, frequency and frequency derivatives.
Table~\ref{t:parameters} presents the best-fit timing solutions.

For two of the pulsars, PSRs \JpsrB{} and \JpsrC{}, the timing
analysis reveals the presence of glitches, manifested as abrupt
changes of the stars' rotation rates.  These significantly complicate
the timing procedure: if not additionally accounted for, they can lead to
loss of phase-coherence.

\begin{figure*}[t]
\centering
		(a) \psrB{}\\[-0.3cm]
		\subfigure
		{\includegraphics[width=1.75\columnwidth]{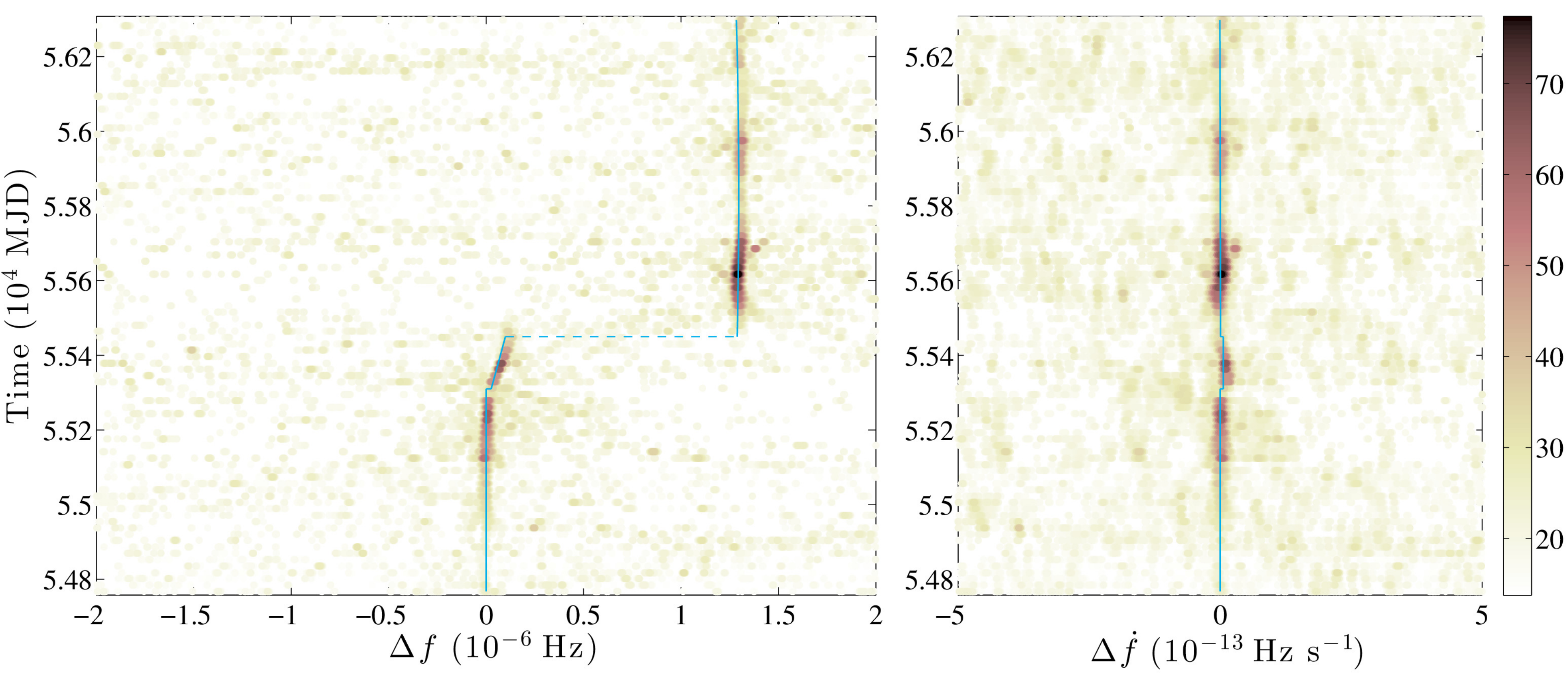}}
		\vspace{0.5cm}\\
		(b) \psrC{}\\[-0.3cm]
		\subfigure
		{\includegraphics[width=1.75\columnwidth]{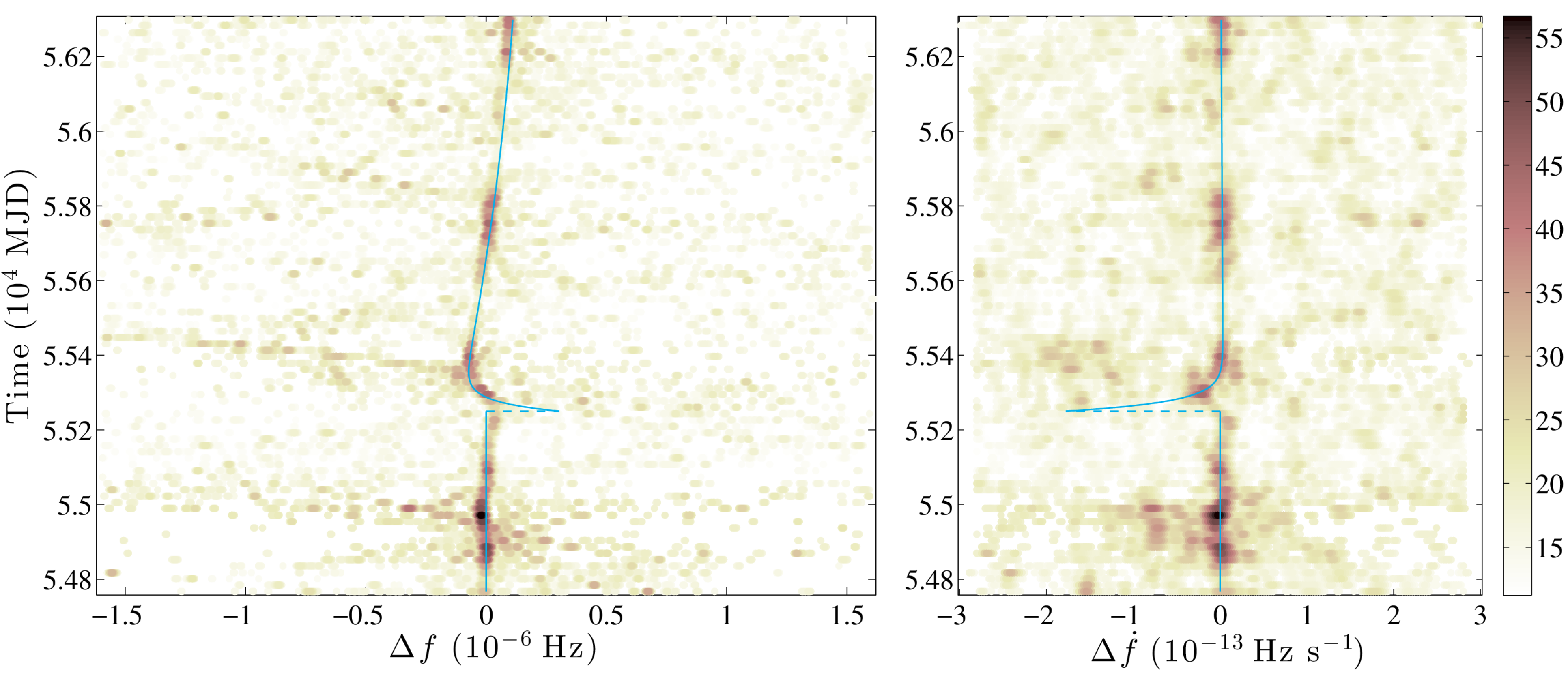}}
		\caption{
Pulsar glitch analyses for \psrB{} (a) and \psrC{} (b). 
The color code represents the weighted $H$-test using photons
within a $170$-day time window that is slid over the entire data set, 
with 90\% overlap.
At fixed (pre-glitch) sky position, 
for each window scans in $H$-test over $\{f,\dot f\}$ are done.  
Vertical axes show the time midpoint of each window. 
Horizontal axes show the offsets from the pre-glitch
parameters in $f$ (left), $\dot f$ (right).
The solid blue curves are superimposed to show the timing solutions 
of Table~\ref{t:parameters}. 
\label{f:glitches}
}
\end{figure*}

In dedicated studies, we examined the spin-parameter changes associated with 
the glitches. For each of the two pulsars, 
we fixed the sky position to the pre-glitch timing solution, and scanned ranges 
in $\{f,\dot f\}$  on a dense grid around the pre-glitch spin parameters.
At each grid point we computed the weighted $H$-test statistic \citep{KerrWeightedH2011} 
using photons within a fixed 
time window. This window was slid over the entire data set with 90\% overlap
between subsequent steps. The results are shown in Figure~\ref{f:glitches}.
The choice of time-window size balances signal-to-noise ratio and time resolution, 
being just long enough to still accumulate a detectable signal-to-noise ratio.
These results enabled us to estimate values for the spin-parameter changes and 
glitch epochs for PSRs~\JpsrB{} and \JpsrC{}. We then iterated the  
timing procedure including a corresponding glitch model. 
The inferred glitch parameters are given in Table~\ref{t:parameters}.
In addition to permanent changes in $f$ and $\dot f$, the glitch model for \JpsrC{} 
also includes a frequency increment $\Delta f_{\rm d, glitch\,1}$ that decays
exponentially on the timescale~$\tau_{\rm d, glitch\,1}$ \citep{Tempo2}. 
Thus, the net effect after this spin-up glitch recovery is a spin-down, 
as shown in Figure~\ref{f:glitches}(b). 

\begin{figure*}[t]
\centering
		%
		\subfigure[\psrA]
		{\includegraphics[width=4.2cm]{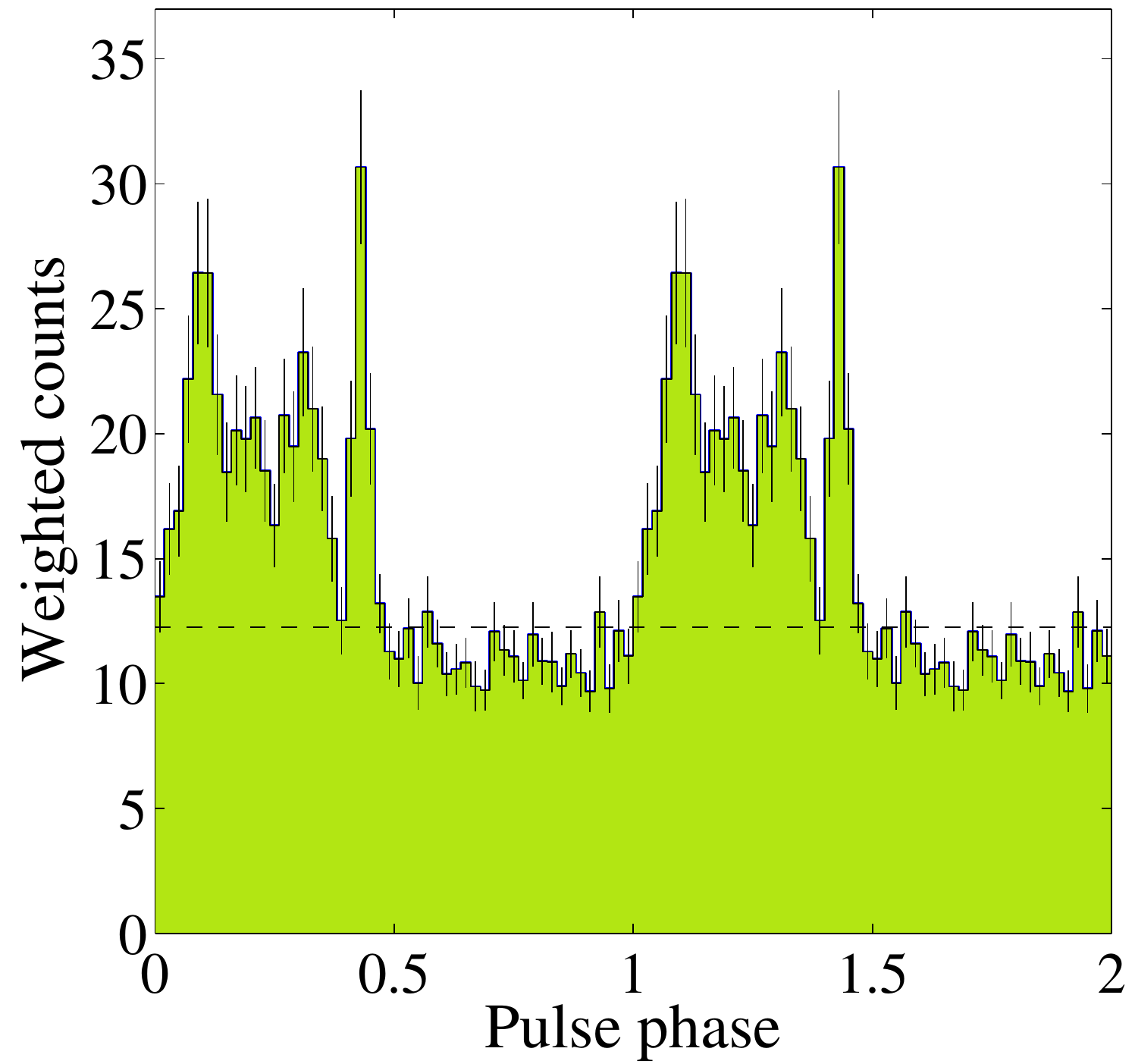}}
		\hspace{0.2cm}
		\subfigure[\psrB]
		{\includegraphics[width=4.2cm]{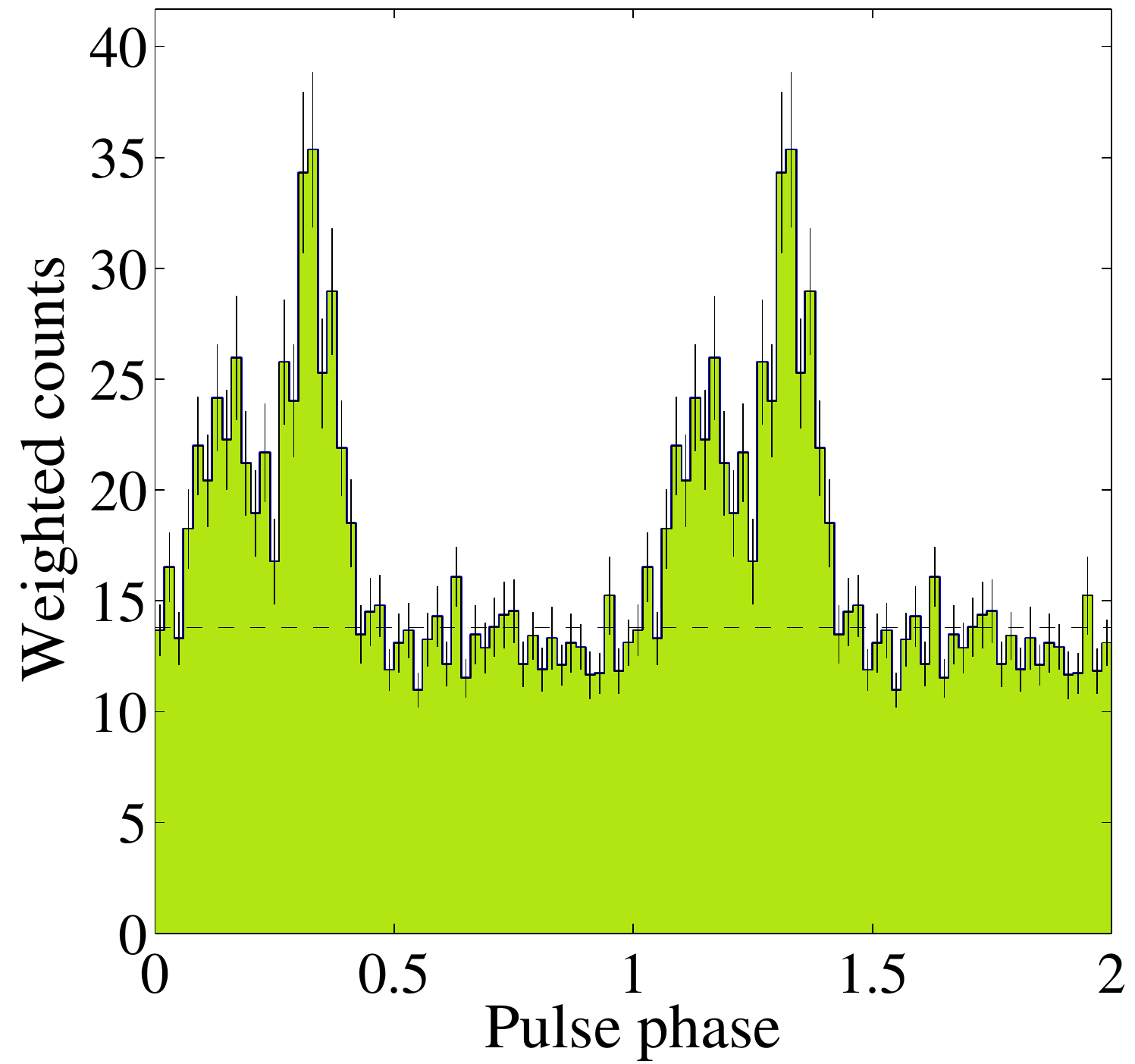}}
		\hspace{0.2cm}
		\subfigure[\psrC]
		{\includegraphics[width=4.2cm]{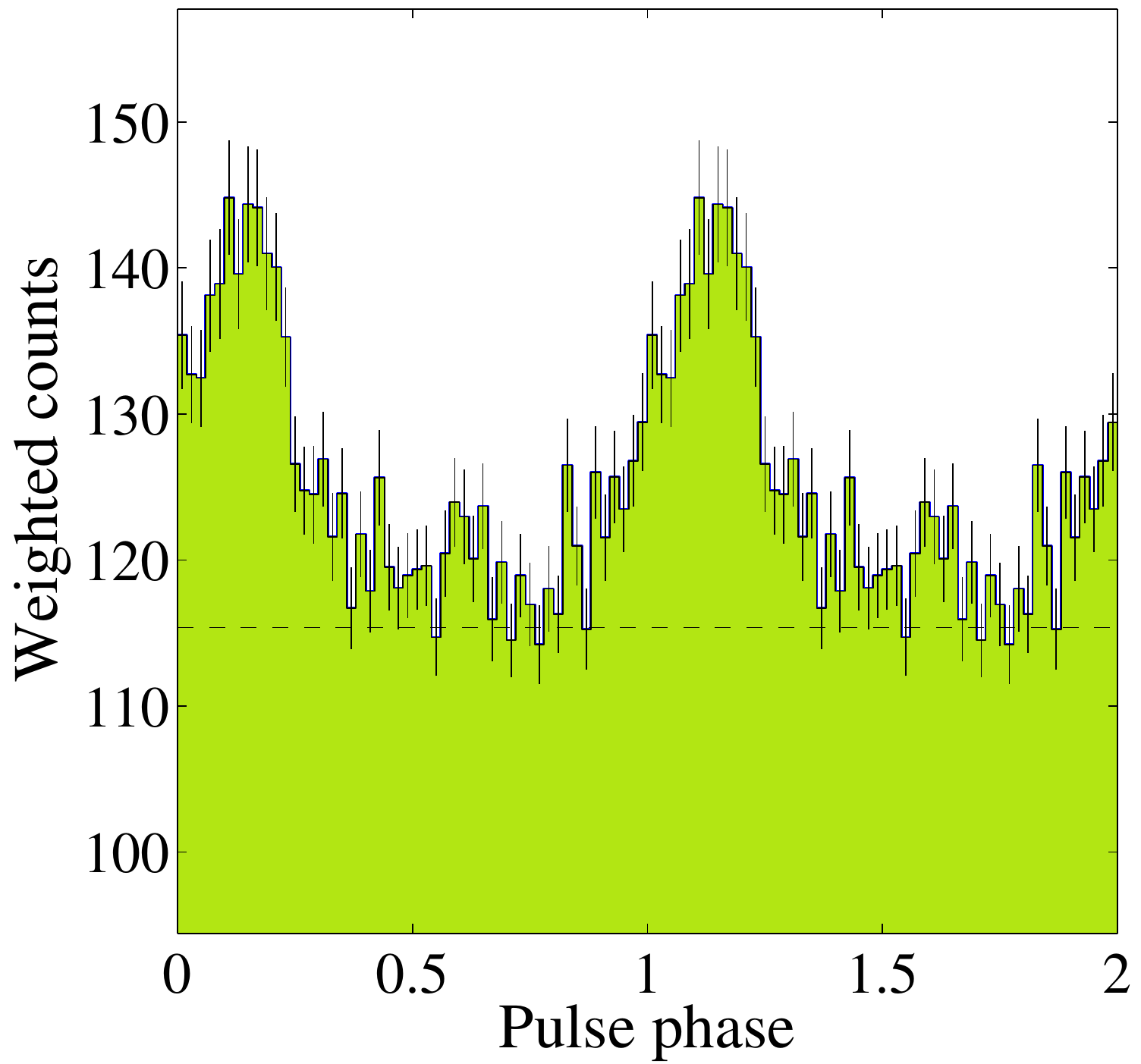}}
		\hspace{0.2cm}
		\subfigure[\psrD]
		{\includegraphics[width=4.2cm]{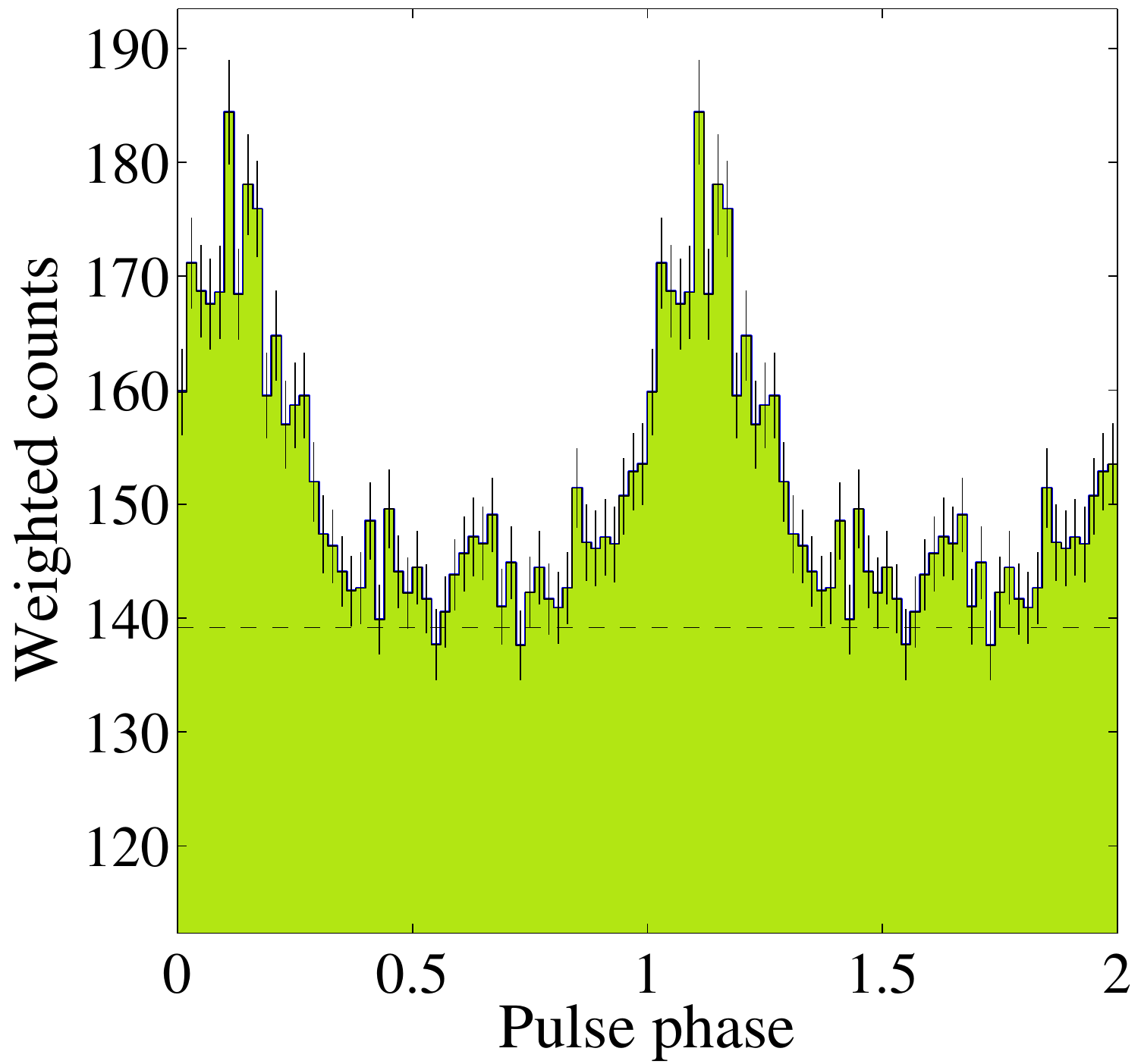}}\\	
		\hspace{-0.2cm}
		\subfigure
		{\includegraphics[width=4.3cm]{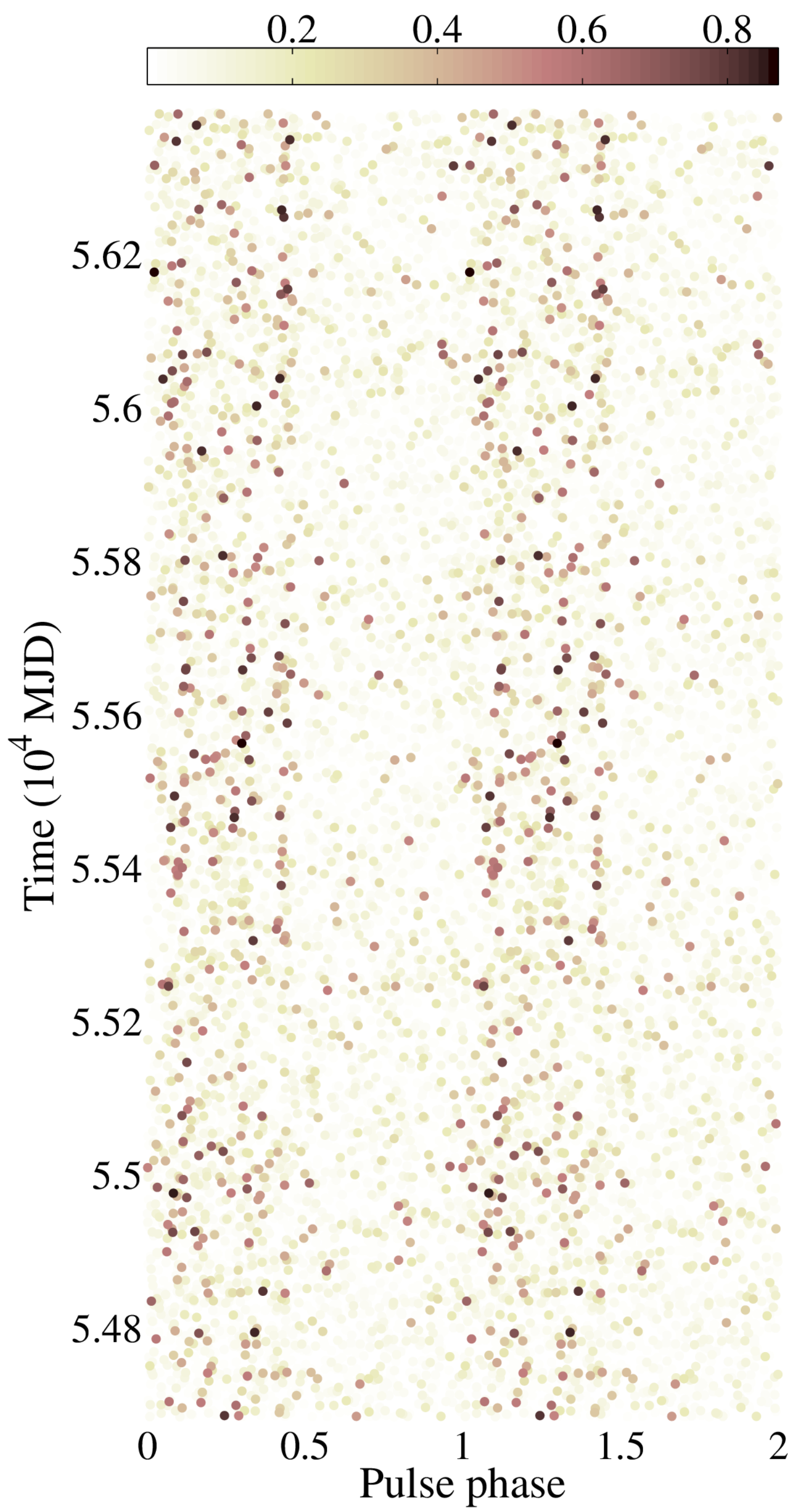}}
		\hspace{0.1cm}
		\subfigure
		{\includegraphics[width=4.3cm]{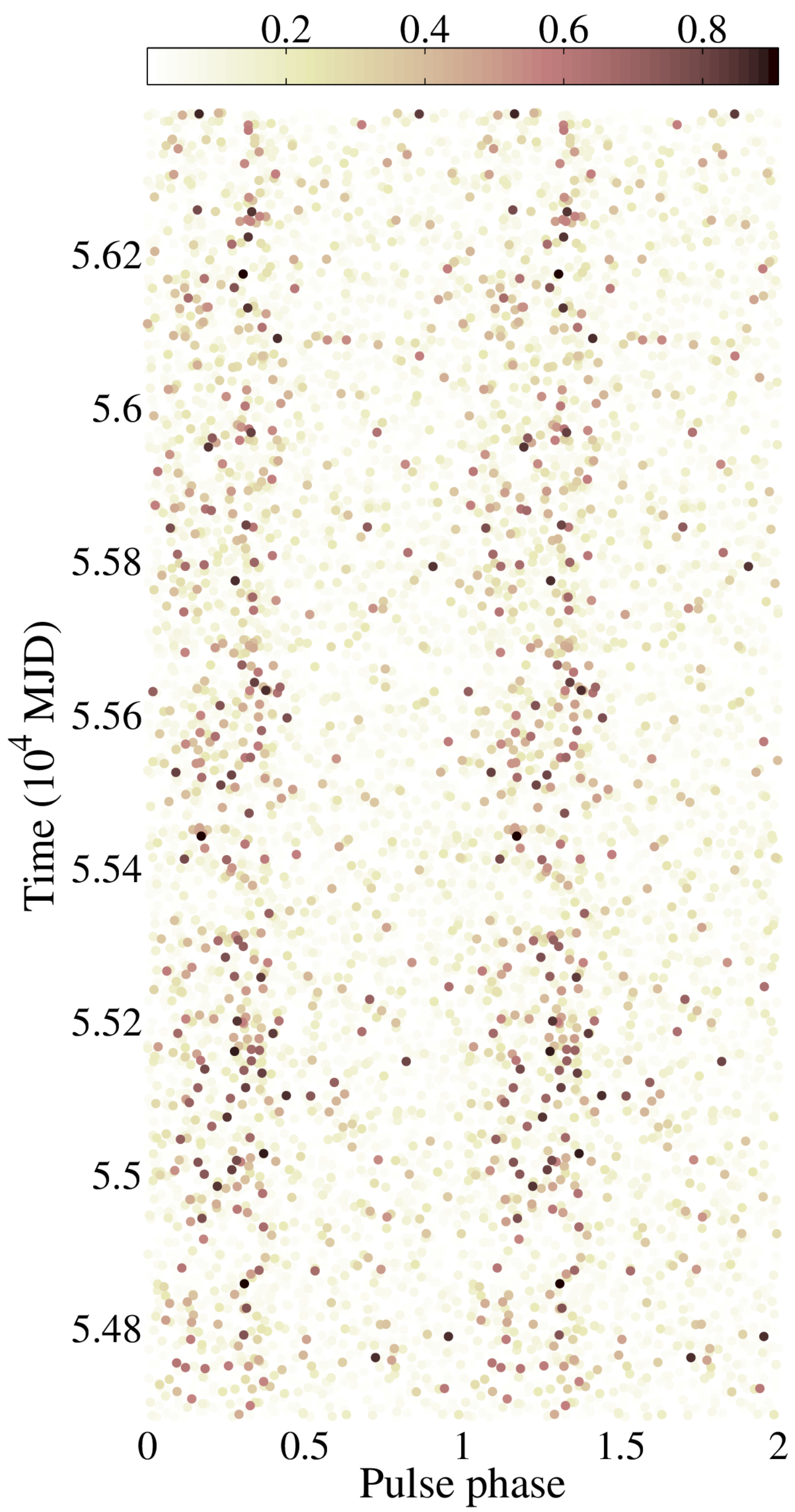}}
		\hspace{0.1cm}
		\subfigure
		{\includegraphics[width=4.3cm]{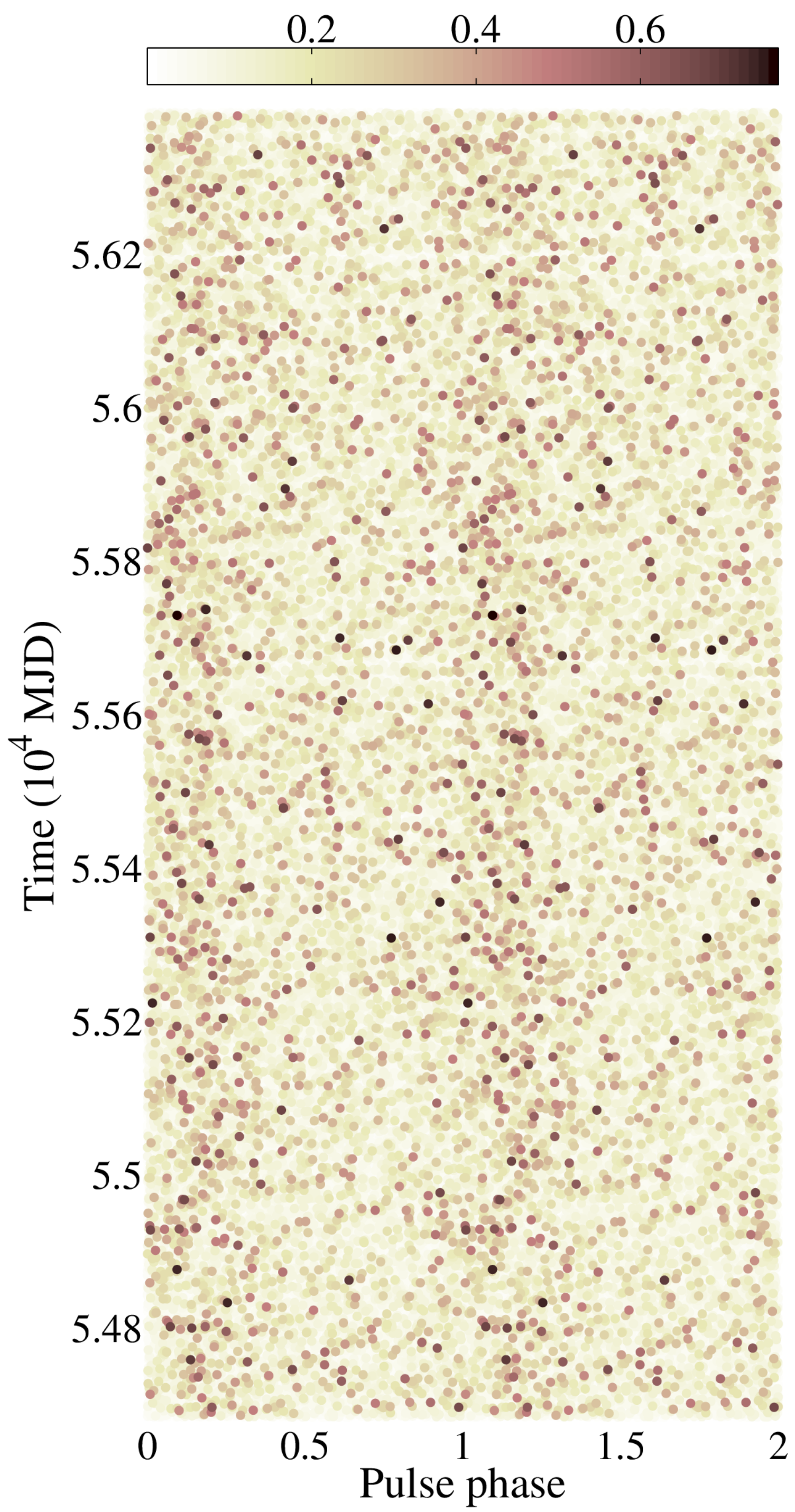}}
		\hspace{0.1cm}
		\subfigure
		{\includegraphics[width=4.3cm]{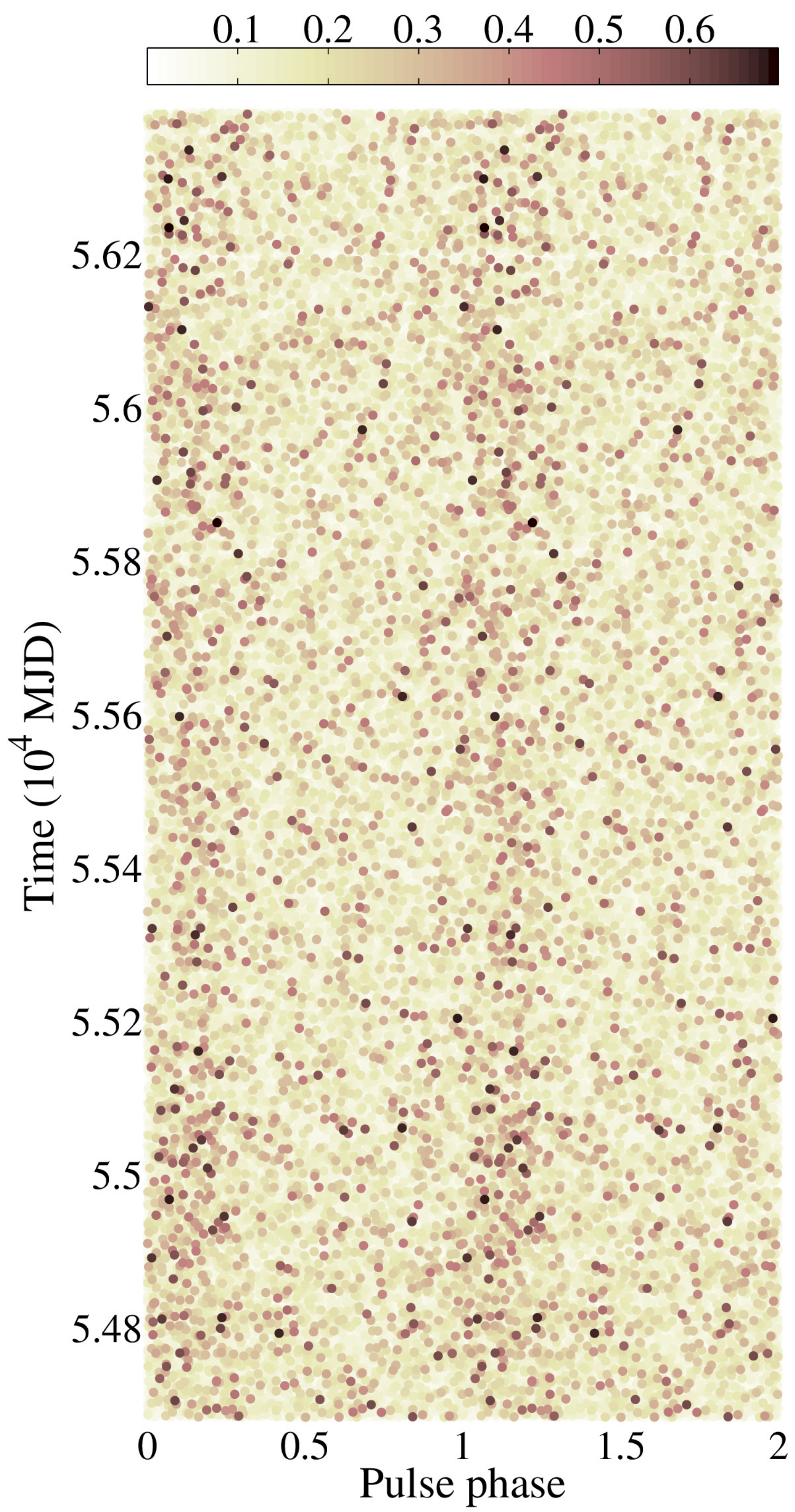}}
	\caption{
	Top row: Gamma-ray pulse profiles using the probability weights 
	 for photons with energies above $100$\,MeV. 
         Each bin is 0.02 in rotation phase.
         Error bars represent 1$\sigma$ statistical uncertainties.
         The dashed line indicates the estimated background level 
         computed from the weights as in \citet{Guillemot2012}.
         Each plot shows two pulsar rotations for clarity.
         Absolute phase references are arbitrarily set to $0.1$ for 
         the first peak.
        Bottom row: The pulsar rotational phase for each gamma-ray-photon 
        arrival time; photon weights are shown in color code.                   
	\label{f:ph-vs-t-all}}
\end{figure*}

Figure~\ref{f:ph-vs-t-all} shows the integrated pulse profiles and 
phase-time diagrams obtained from the full timing solutions.
To characterize the profiles we fitted the observed gamma-ray light curves 
with combinations of Lorentzian and/or Gaussian lines. 
The complex pulse profile of \psrA{} was fitted using asymmetric Lorentzian 
lines for the first two peaks, and a simple Gaussian for the last component. 
For \psrB{} the best fit is based on two simple Gaussian lines. 
For PSRs \JpsrC{} and \JpsrD{} we modeled the light curves 
with asymmetric Lorentzian lines.  Table~\ref{t:parameters} shows 
the resulting peak separations and FWHMs.

We measured the pulsars' phase-averaged spectral properties 
through a binned likelihood analysis, using \textit{pyLikelihood} of the STs. 
We constructed spectral models including all sources found within 20\degr of the pulsars 
from an internal catalog of gamma-ray sources based on three years of LAT data,
where the parameters only of point sources within 5\degr were left free.
Each pulsar spectrum was modeled as an exponentially cutoff power law, 
$dN / dE \propto E^{-\Gamma} \exp\left( - E / E_c \right)$, 
where $\Gamma$ denotes the spectral index and $E_c$ is the cutoff energy. 
The source models included contributions 
from the Galactic diffuse emission (using model \textit{gal\_2yearp7v6\_v0}), 
the extragalactic diffuse emission, 
and the residual instrumental background (using template \textit{iso\_p7v6source} \footnote{\href{http://fermi.gsfc.nasa.gov/ssc/data/access/lat/BackgroundModels.html}{http://fermi.gsfc.nasa.gov/ssc/data/access/lat/BackgroundModels.html}}). 
For \psrB{}, the phase-averaged analysis could not constrain $\Gamma$.
Excluding ``off-pulse'' photons with phases between $0.55$ and $0.95$ 
(Figure~\ref{f:ph-vs-t-all}) slightly improved the fit quality. 
The best-fit values for $\Gamma$, $E_c$, and the 
derived photon and energy fluxes are given in Table~\ref{t:parameters}. 
These are in line with $\Gamma$ and $E_c$ values of other young 
LAT pulsars \citep{Fermi2PC}, apart from \psrB's low $\Gamma$
which is currently not well constrained. 
Future LAT-event-reconstruction enhancements \citep[][]{Atwood+2013} and 
more photon data may improve the latter measurement.

\section{Radio Counterpart Searches}\label{s:mw}

We searched for radio pulsations from the new gamma-ray pulsars 
with the Effelsberg and Parkes Telescopes. 
PSRs \JpsrA{} and \JpsrD{} were observed with the Effelsberg Telescope 
at $1.4$\,GHz for $1$\,hr, using the new Ultra Broad Band (UBB) 
receiver (R. Keller et al. 2013, in preparation), as part of the commissioning of the instrument. 
For \psrB{}, we used four $72$-minute pointings taken during
the High Time Resolution Universe survey \citep{Keith+2010}, 
which we combined and searched coherently for increased sensitivity. 
Finally, for \psrC{} we analyzed a $2.4$-hr Parkes observation.
We used the gamma-ray pulsar ephemerides (Table~\ref{t:parameters}),
valid for each radio observation, to fold the radio data
and searched in dispersion measure up to $2000$\,pc~cm$^{-3}$. 

We found no significant detection and provide upper limits on the radio 
flux density, derived with the modified radiometer equation \citep{psrhandbook} 
assuming a detection signal-to-noise-ratio threshold of $5$ and a $10$\% duty cycle.
Configuration details for the Parkes observations 
are found in Table~5 of \citet{Pletsch+2012-9pulsars}. 
The observation of \psrC{} pointed 1.5\arcmin{} away from the pulsar.
This implies only a small sensitivity loss, because the beam's half-width at half-maximum 
is 7\arcmin{} at this frequency. We accounted for this loss as in \citet{Pletsch+2012-9pulsars}. 
For the $1.4$\,GHz Effelsberg observations, we assumed $n_p = 2$, $\beta = 1.2$, and 
a system equivalent flux density $T_\mathrm{sys} / G = 40.5$ Jy, 
as measured from preliminary UBB performance estimates\footnote{UBB performance estimates may evolve 
as commissioning continues.} from
continuum observations of 3C~48. After removal of radio-frequency interference, 
the nominal frequency bandwidth of $260$\,MHz was reduced to $225$ and $205$\,MHz 
for PSRs \JpsrA{} and \JpsrD{}, respectively. 
Table~\ref{t:parameters} lists the resulting radio-flux-density limits $S_{1400}$ at $1.4$\,GHz, 
which are at the higher end compared to other LAT-discovered pulsars \citep[e.g.,][]{Fermi2PC}.
Although, assuming ``heuristic'' distances $d_{h}$ (Table~\ref{t:parameters}), the
luminosities $S_{1400}\, d_{h}^2$ are lower than for the vast majority of known pulsars,
deeper radio searches are possibly warranted to confirm the present picture.

\section{Discussion}\label{s:discussion}

The measured spin parameters of all four new gamma-ray pulsars
classify them as young and energetic (Figure~\ref{f:f-fdot-diagr}).
Their spin-down powers, $\dot E = -4\pi^2 I f \dot{f}$, 
range from $5.6\times10^{34}$ to $1.2\times10^{36}$ erg s$^{-1}$,
for an assumed neutron-star moment of inertia of $I=10^{45}$~g~cm$^2$.
With characteristic ages $\tau_c$ between $35$ and $56$~kyr, 
they are among the youngest 4\% of pulsars 
known~\citep{ATNFcat}\footnote{\href{http://www.atnf.csiro.au/research/pulsar/psrcat/}{http://www.atnf.csiro.au/research/pulsar/psrcat/}}. 

The distances to the four new objects are difficult to constrain
without detected radio pulsations providing a dispersion measure.
However, an estimated upper bound for the distance~$d$ to each 
pulsar can be obtained from relating
 $\dot E$ and the gamma-ray luminosity, $L_\gamma=4\pi\,d^2\,f_\Omega\,G_{100}$,
where $f_\Omega$ is a beam correction factor \citep{Watters2009}.
Assuming 100\% conversion efficiency ($L_\gamma = \dot E$) 
as an upper limit, and $f_\Omega\sim1$, typical of gamma-ray 
pulsars \citep{Watters2009}, the above relation can then be solved for distance,
which we denote by $d_{100\%}$. The resulting $d_{100\%}$ upper limits 
for the four pulsars are between $5$ and $12$\,kpc.
More realistically, if instead a ``heuristic'' gamma-ray luminosity \citep[as in][]{Fermi2PC},
\mbox{$L_\gamma^h = (\dot E/10^{33}$\,erg s$^{-1})^{1/2}$\,$10^{33}$\,erg s$^{-1}$} is assumed, 
distances between $1.5$ and $2.1$\,kpc result (Table~\ref{t:parameters}),
suggesting that the pulsars are rather close.

Pulsars are believed to form in supernovae, so the discovery of a young pulsar
prompts us to look for an associated supernova remnant (SNR).
From \citet{GreenSNRcat}, we find a co-located SNR for
two of the newly discovered pulsars. 
Care should be taken to establish genuine pulsar/SNR associations,
because a chance superposition on the sky has a non-negligible 
probability~\citep[e.g.,][]{Gaensler+1995,Kaspi1998}.

The sky position of \psrA{} lies about $6\arcmin$ from the geometric center of SNR \snrA{},
which has an angular size of $70\arcmin$ \citep{Fuerst+1986}. 
The low surface brightness $\Sigma$ reported by \citet{Fuerst+1986}  
suggests a large SNR age of $10$--$100$\,kyr, which 
is compatible with the pulsar's characteristic age $\tau_c = 52$\,kyr. This
is a good estimator of the pulsar's true age if (1) the present spin period ($P=465$\,ms) is much 
larger than at birth and (2) the spin-down is dominated by magnetic dipole braking. 
From the $\Sigma$-diameter relation \citep{Milne1979}, an SNR size of $\approx70$\,pc yields 
a distance of $\approx3.5$\,kpc. This has a large uncertainty, but is
compatible with our estimated maximum pulsar distance
$d_{100\%}=5.2$\,kpc.
Depending on the estimated age and distance of the SNR, the required
transverse velocity of \psrA{} is between $60$ and $850$\,km~s$^{-1}$,
which is within the typical range of other pulsars \citep[e.g.,][]{Hobbs+2005}.
Thus it appears plausible that \psrA{} and \snrA{} are associated.

The $102$\,ms pulsar \JpsrC{} is located about $11\arcmin$ from the centroid of SNR \snrC{}, 
which has an extension of $31\arcmin \times 23\arcmin$. \citet{Caswell+1975}
provide an estimated distance of $5.5$\,kpc, compatible with
our $d_{100\%}=12.5$\,kpc estimated upper limit for \psrC{}.
Their SNR age estimate of $20$--$100$\,kyr is also compatible with the
$\tau_c = 52$\,kyr characteristic age of \psrC{}, given the same caveats as above.
From the possible values of the estimated SNR age and distance, the necessary
transverse speed of \psrC{} is between $170$ and $860$\,km~s$^{-1}$,
which is also reasonable.
Thus, a genuine association between \psrC{} and \snrC{}
is plausible and merits further study.

\section{Conclusion}\label{s:conclusion}

We have reported the \EAH{} discovery and follow-up study of four gamma-ray pulsars  
found in a novel blind-search effort using \Fermi-LAT data. 
The inferred parameters characterize the pulsars as energetic and young,
likely relatively nearby. Young neutron stars are rare, and nearby ones in particular
\citep[e.g.,][]{Keane+2008}. 
As such, these four discoveries contribute
toward a more complete understanding of the young pulsar population and
neutron-star birthrates \citep{Watters+2011}. 
For two of the new pulsars, we have shown that associations with positionally coincident SNRs
are possible. However, confirmation requires further work (e.g., measuring pulsar proper motion, 
a difficult task using LAT data alone).

All four gamma-ray pulsars lie close to the Galactic plane and remained 
undetected in subsequent targeted radio searches. 
In part, this is not unexpected, as argued by \cite{Camilo2012}, 
since the vast majority of Galactic-plane (non-MSP) radio pulsars
detectable by current radio telescopes and producing  
gamma-ray fluxes observable at Earth are likely already known.
In turn, this demonstrates the importance of continued blind pulsar searches  
of gamma-ray data: it is the only way to discover such neutron stars. 
It is also remarkable that PSRs~\JpsrB{} and \JpsrC{} have been detected in the 
blind search despite their prominent glitch activity.
These facts, plus the combination of improved search techniques and 
massive \EAH{} computing power leaves us optimistic that we can find 
more pulsars among the LAT unidentified sources.
\\

\acknowledgements

We thank all \EAH{} volunteers, and especially
those whose computers detected the pulsars with the highest significance:
\psrA: ``David Z'', Canada and ``Test'', France;
\psrB: Thomas M. Jackson, Kentucky, USA and ``mak-ino'', Japan;
\psrC: NEMO computing cluster, UW-Milwaukee, USA and ``Chen'', USA;
\psrD: Doug Lean, Australia and Hans-Peter Tobler, Germany.

This work was supported by the Max-Planck-Gesellschaft (MPG).
The \Fermi{} LAT Collaboration acknowledges support from several agencies and institutes for both development and the operation of the LAT as well as scientific data analysis. These include NASA and DOE in the United States, CEA/Irfu and IN2P3/CNRS in France, ASI and INFN in Italy, MEXT, KEK, and JAXA in Japan, and the K.~A.~Wallenberg Foundation, the Swedish Research Council and the National Space Board in Sweden. Additional support from INAF in Italy and CNES in France for science analysis during the operations phase is also gratefully acknowledged. The UBB receiver construction was supported by the ERC under contract No. 279702.  \EAH{} is supported by the MPG and by US National Science Foundation grants 1104902 and 1105572.

\bibliographystyle{apj}

\bibliography{ms4psrs} 

\end{document}